\documentclass[aps,prl,twocolumn,groupedaddress,10pt]{revtex4-1}

\usepackage{graphicx, subfigure}
\usepackage[utf8]{inputenc}
\usepackage{grffile}
\usepackage{import}
\usepackage{enumerate}

\usepackage{amsfonts}
\usepackage{amsmath}
\usepackage{amssymb}
\usepackage{siunitx}
\usepackage{color}
\usepackage[normalem]{ulem}
\usepackage{todonotes}
\usepackage[bookmarks, 
            breaklinks, 
            pdfstartview=FitH, % fit width when the document is opened
            pdftitle={},
            pdfauthor={},
            pdfkeywords={}
            ]{hyperref}
\usepackage{ulem}

\begin{document}

\title{Interferometry with non-classical motional states of a Bose-Einstein condensate}

\author{S.~van~Frank$^1$}
\author{A.~Negretti$^{2,3}$}
\author{T.~Berrada$^1$}
\author{R.~B\"{u}cker$^{1,4}$}
\author{S.~Montangero$^3$}
\author{J.-F.~Schaff$^1$}
%\email[]{jschaff@ati.ac.at}
\author{T.~Schumm$^1$}
\author{T.~Calarco$^3$}
\author{J.~Schmiedmayer$^{1 \ast}$}
\affiliation{\small $^1$ Vienna Center for Quantum Science and Technology, Atominstitut, TU Wien, Stadionallee 2, A-1020 Vienna}
\affiliation{\small $^2$ Zentrum f\"ur Optische Quantentechnologien, Universit\"at Hamburg, The Hamburg Centre for Ultrafast Imaging, Luruper Chaussee 149, D-22761 Hamburg}
\affiliation{\small $^3$ Institut f\"ur Quanteninformationsverarbeitung, Universit\"at Ulm, Albert-Einstein-Allee 11, D-89069 Ulm}
\affiliation{\small $^4$ Max Planck Institute for the Structure and Dynamics of Matter, Bldg. 99 (CFEL), Luruper Chausse 149, D-22761 Hamburg}

\date{\today}

%----------------------------------------------------------------------------------------
%	INTRODUCTORY PARAGRAPH
%----------------------------------------------------------------------------------------

\begin{abstract}
\textbf{We demonstrate a two-pulse Ramsey-type interferometer for non-classical motional states of a Bose-Einstein condensate in an anharmonic trap. The control pulses used to manipulate the condensate wavefunction are obtained from Optimal Control Theory and directly optimised to maximise the interferometric contrast. They permit a fast manipulation of the atomic ensemble compared to the intrinsic decay and many-body dephasing effects, and thus to reach an interferometric contrast of \SI{92}{\percent} in the experimental implementation.}
\end{abstract}

%----------------------------------------------------------------------------------------
%	ARTICLE CONTENT
%----------------------------------------------------------------------------------------

\maketitle

The control of quantum states is an important building block for fundamental investigations and technological applications of quantum physics~\cite{Mabuchi2005}. However, quantum many-body systems like Bose-Einstein condensates (BEC) exhibit complex behaviours that make them difficult to manipulate~\cite{Bloch2008}, in particular in the presence of intrinsic dephasing~\cite{Gring2012,Berrada2013b}, decoherence or decay~\cite{Bucker2011}. One strategy to control such quantum states is to implement operations faster than the characteristic timescales of the prejudicial processes, using for example optimal control theory (OCT)~\cite{Brif2010,DAlessandro2007}. The speedup can be exploited to realise elaborate manipulations, in the present case a sequence of state transfer pulses for interferometry.

The Ramsey interferometer~\cite{Ramsey1990} is a prime example of precise control of a system at the quantum level. It is usually implemented using internal states of atoms~\cite{Guena2012}, molecules~\cite{Ramsey1956} or ions~\cite{Monroe1996}, for which powerful manipulation procedures are now available. In contrast, controlling a more complex many-body system to this level remains a challenge.

\begin{figure*}[t!]
	\centering
	\def\svgwidth{1.8\columnwidth}
	%\import{../figures/}{Figure1_Schematics.pdf_tex}
	\includegraphics[width=0.9\linewidth]{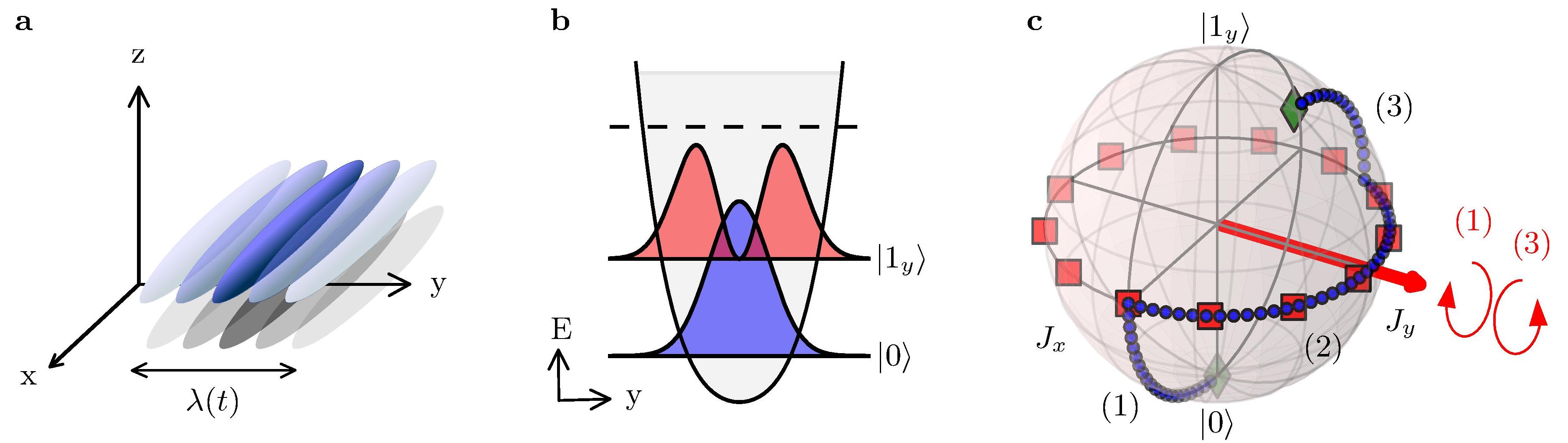}
	\caption{\textbf{Schematic of the Ramsey interferometric sequence}. \textbf{(a)} Representation of the quasi-BEC subjected to a fast displacement $\lambda(t)$ in the $y$-direction. \textbf{(b)} Trapping potential and effective two-mode system. The anharmonicity  in the $y$-direction leads to a unique transition frequency between the ground state $|0\rangle$ and the lowest-lying excited state $|1_y\rangle$, effectively almost isolating the two-level system $|0\rangle$ - $|1_y\rangle$. The other states (dashed line) have higher energies. \textbf{(c)} Example of an interferometric trajectory (blue dots) on the Bloch sphere representation of the two-level system. (1) is the first $\pi / 2$ pulse that prepares a balanced coherent superposition. (2) is the phase accumulation time corresponding to a rotation around the vertical axis. (3) is the second $\pi / 2$ and corresponds to a $\SI{90}{\degree}$ counter-clockwise rotation around $J_y$. The red squares show the 15 points on which the second $\pi / 2$ pulse was optimised (see Supplementary Information). }
	\label{fig:figure1}
\end{figure*}

In this letter, we report the implementation of a Ramsey interferometer sequence for the motional states of a Bose-Einstein condensate in a trap. Such an interferometer requires the following operations: (i) the creation of a superposition between two trap eigenstates and (ii) a general ``$\pi / 2$'' pulse independent of the input state for the read out (Fig. \ref{fig:figure1}(c)). Transitions between motional states are driven by ``shaking'' the trap along one axis (Fig.~\ref{fig:figure1}(a)). The space of accessible motional states is almost reduced to an effective 2-level system by making the trapping potential slightly anharmonic along one direction~\cite{Bucker2013} (Fig.~\ref{fig:figure1}(b)). Combined with the use of OCT pulses, this limits the probability of leakage to higher states. 

We design the two control pulses using the chopped random basis algorithm (CRAB)~\cite{Caneva2011}, with the particularity that the second pulse is directly optimised to reach high interferometric contrast. For this optimisation, we describe the system's dynamics as a condensate wavefunction using an effective one-dimensional Gross-Pitaevskii equation (1D GPE), which is justified by the very low temperatures and the very short times needed. The obtained pulses allow us to drive transitions between the specified motional states~\cite{Bucker2013} at a timescale comparable to the trapping frequency~\cite{Bason2011}. The produced states are then superpositions of two (or few) motional Fock states~\cite{Bouchoule1999, Monroe1996}, to be distinguished from the Poissonian superpositions of motional states populated in a classical center-of-mass movement.

% ---- Figure 1 ----

%----------------------------------------------
%\section{Experimental procedure}

\begin{figure}[b!]
	\centering
	\def\svgwidth{\columnwidth}
	%\import{../figures/}{Figure2_First_pulse.pdf_tex}
	\includegraphics[width=\linewidth]{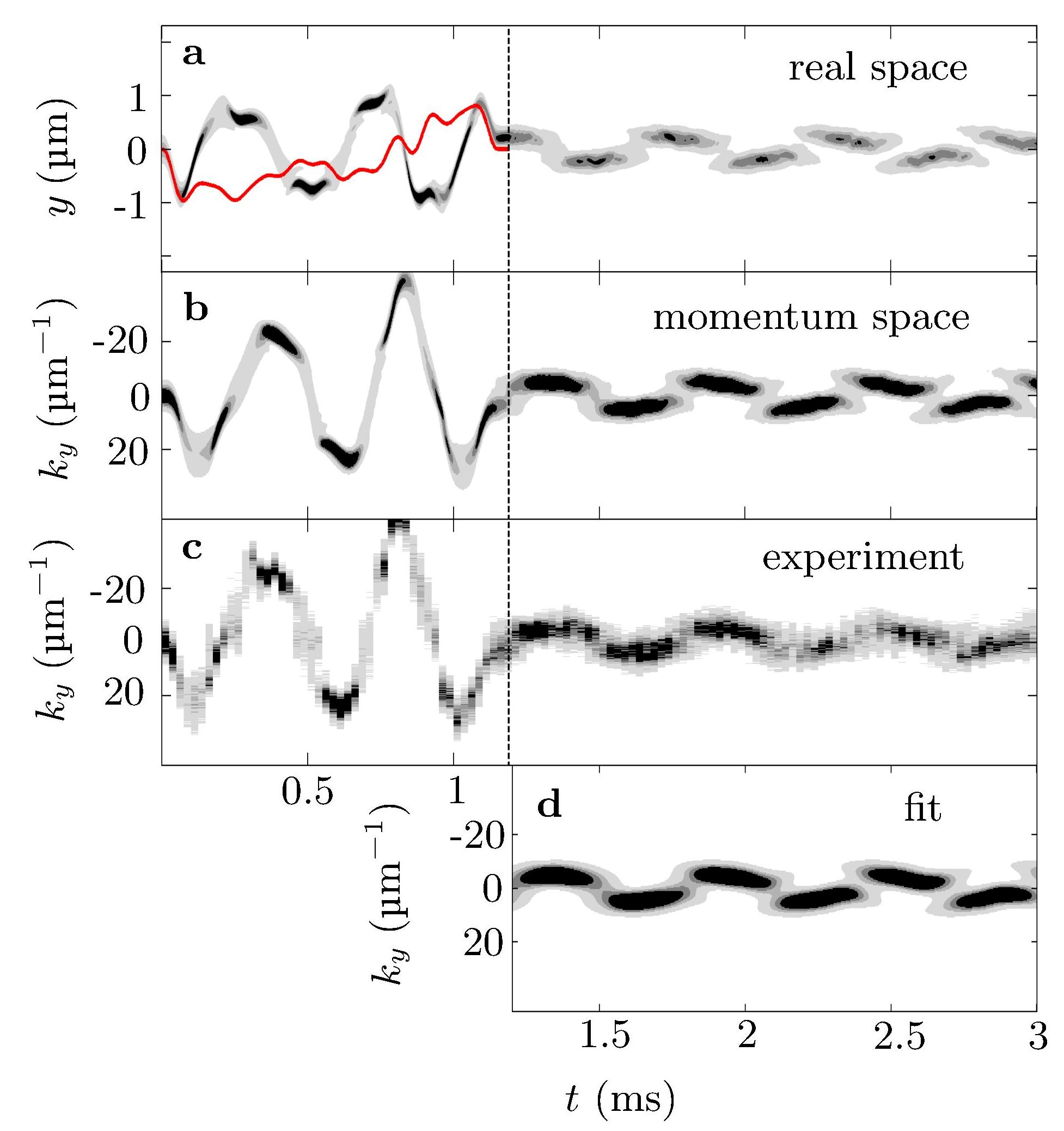}
	\caption{\textbf{Dynamics of the excitation and interference patterns observed during and after the first $\pi/2$ pulse. (a)} Optimised trap displacement $\lambda(t)$ along the $y$-direction (red solid line) and simulation of the in-situ density. \textbf{(b)} Simulated momentum distribution. \textbf{(c)} Measured momentum distribution. The time-of-flight images were integrated along the longitudinal $x$-direction and concatenated to show the time evolution. \textbf{(d)} Fit to the momentum distribution from which the populations $p_0$ and $p_1$ are extracted (see text).}
	\label{fig:figure2}
\end{figure}

Our experimental system, sketched in Fig.~\ref{fig:figure1}, is a dilute, quasi one-dimensional quantum degenerate gas of $\sim 700$ $^{87}$Rb atoms in an elongated magnetic trap on an atom chip~\cite{Reichel2011}. Both the temperature $T < \SI{50}{\nano\kelvin}$ and chemical potential $\mu/h \simeq \SI{0.6}{\kilo\hertz}$ are below the smallest transverse level spacing $ E_{01} = h \times \SI{1.83}{\kilo\hertz}$, ensuring that the system is initialised in the motional ground state $|0\rangle$ (see Supplementary Information). Radio-frequency dressing introduces an anharmonicity in the horizontal transverse $y$-direction (see Supplementary Information). We drive transitions between the two lowest-lying motional states by displacing the trap purely along the $y$-direction, following trajectories obtained by the CRAB optimisation. The trap displacement $\lambda(t)$ reaches values on the order of 4 times the rms size of the ground state wavefunction (see Fig.~\ref{fig:figure2}(a)).

%----------------------------------------------
%\section{State analysis}

The behaviour of the wavefunction in the horizontal $xy$-plane at different times $t$ throughout the Ramsey sequence is monitored by time-of-flight fluorescence imaging~\cite{Bucker2009}. Along the transverse $y$-axis, the high trap frequency and $\SI{46}{\milli\second}$ expansion time ensure that the measured atomic density is an image of the in-trap momentum distribution. The experimental images are integrated along the longitudinal $x$-axis and concatenated to follow the evolution of the transverse wavefunction over time, as illustrated in Fig.~\ref{fig:figure2}(c).

After the control pulses, the density distributions exhibit characteristic ``beating'' patterns arising from interferences between the different motional levels populated. Comparing the time-dependent momentum distribution to the GPE simulations allows us to extract the GPE eigenstate populations. From this, we estimate the fidelity of the first $\pi/2$ pulse as well as the output of the full interferometric sequence. (see Supplementary Material and~\cite{Bucker2013})

%----------------------------------------------
%\paragraph{First $\pi / 2$ pulse:}

The first $\pi / 2$ pulse (Fig.~\ref{fig:figure2}(a)) aims to create a balanced superposition \mbox{$| \psi_{\mathrm{target}}\rangle=\frac{1}{\sqrt{2}}(|0\rangle+e^{i \phi}|1_y\rangle)$} of the ground state $|0\rangle$ and first excited state $|1_y\rangle$ with a relative phase $\phi$ arbitrarily chosen to be zero. The cost function to be minimised can be written in terms of the overlap fidelity $\mathcal{F}$:
\begin{align}
\label{eq:fidelity}
\mathcal{J}^{(1)}= 1 - \mathcal{F} = 1 - \Re\left[
\langle\psi_{\mathrm{target}}| \psi(T_{\pi/2}^{(1)}) \rangle
\right]^2,
\end{align}
where $| \psi(T_{\pi/2}^{(1)}) \rangle$ represents the state of the system at the end of the first $\pi / 2$ pulse. 

When designing the pulse, a trade-off must be found between fidelity and speed \cite{Doria2011,Bason2011}. We choose a pulse with a theoretical fidelity of $\SI{98.6}{\percent}$ for a pulse duration of $\SI{1.19}{\milli\second}$. This duration is about twice the timescale set by the single-particle level spacing $\nu_{01}^{-1}  = h/E_{01} = \SI{0.55}{\milli\second}$.

% ---- Figure 2 ----

%----------------------------------------------
%\paragraph{Phase accumulation time:}

\begin{figure}[t]
	\centering
	\def\svgwidth{\columnwidth}
	%\import{../figures/}{Figure3_Interferometric_fringes.pdf_tex}
	\includegraphics[width=\linewidth]{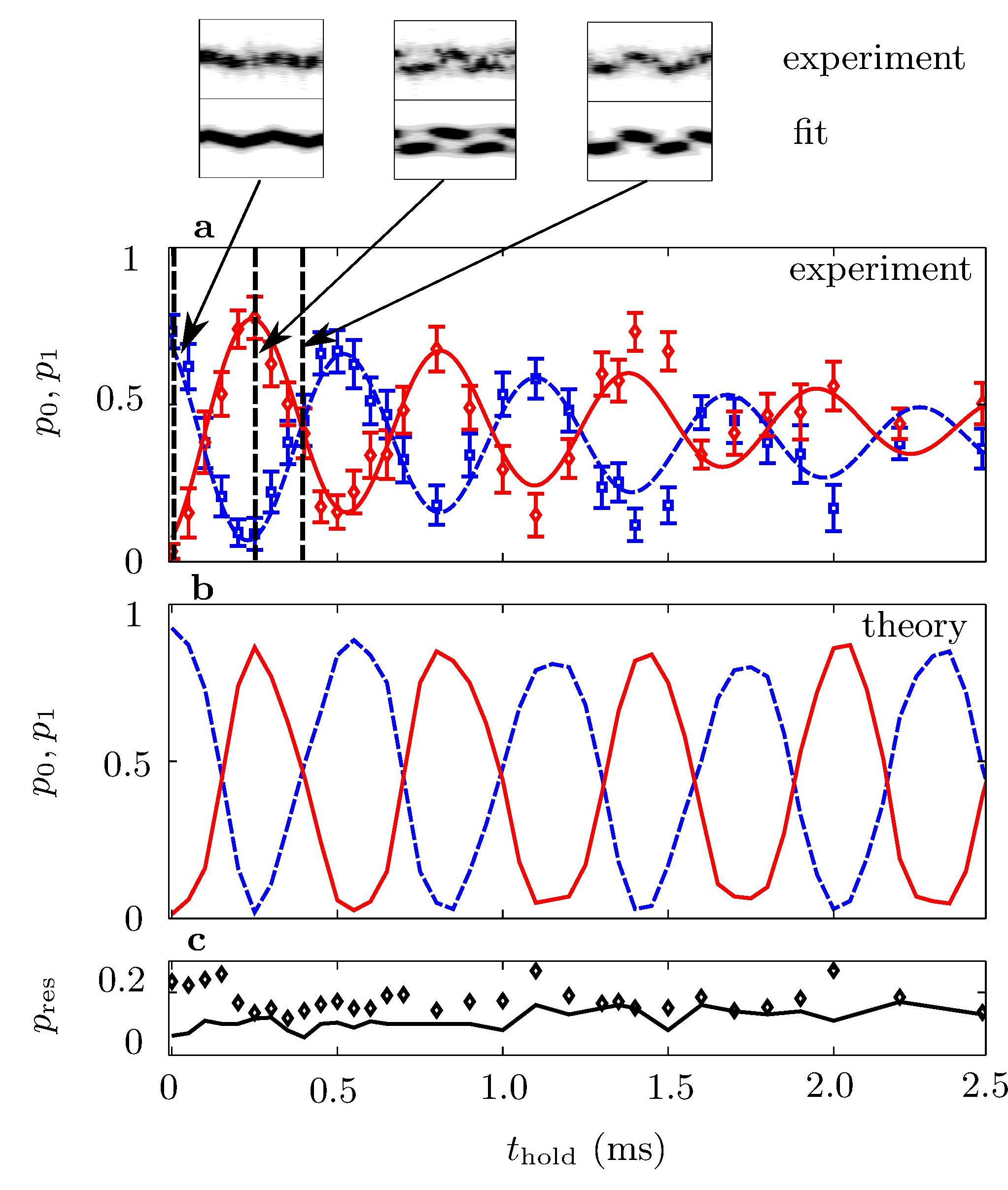}
	\caption{\textbf{Interference fringes of the motional-states interferometer.} \textbf{(a)} Experimental data. Populations of the ground state $p_0$ (blue circles) and first excited state $p_1$ (red diamonds), extracted from a fit to the experimental density images, as a function of the phase accumulation time $t_{\mathrm{hold}}$.  The error bars indicate the $1 \sigma$ confidence interval of the fit. The blue and red dashed lines are exponentially damped sines.  \textbf{(b)} OCT optimisation data. Populations of the ground state $p_0$ (blue dashed line) and first excited state $p_1$ (red line) as a function of the phase accumulation time $t_{\mathrm{hold}}$ \textbf{(c)} Populations in higher excited states in the optimisation (black solid line) compared to residual part in the fits to experimental data (black diamonds). The top insets are examples of experimental momentum distributions (upper) and their corresponding fitted GPE momentum distribution (lower) for the 3 different hold times indicated by the vertical dashed lines in panel (a).} 
	\label{fig:figure3}
\end{figure}

After creating a coherent superposition of $|0\rangle$ and $|1_y\rangle$, the wavefunction is held in a static potential for an adjustable time $t_{\mathrm{hold}}$. 
The energy difference between the levels leads to an evolution of the relative phase. In a simplified linear picture, this phase evolution corresponds to a rotation of the state vector on the equatorial plane of the Bloch sphere at a constant angular frequency given by the energy difference between the levels (see Fig.~\ref{fig:figure1}(c)). In the trap, the inter-atomic interactions introduce a non-linearity in the system and the corresponding mean-field energy gives a small additional contribution on top of the single-particle energy splitting $E_{01}$. For a balanced superposition, one period of the oscillation of the relative phase is then $T = \SI{0.58}{\milli\second}$, namely a \SI{5}{\percent} increase. This phase accumulation time $t_{\mathrm{hold}}$ is varied to observe interferometric fringes in the Ramsey sequence (Fig.~\ref{fig:figure3}).

%----------------------------------------------
%\paragraph{Full Ramsey sequence}

The second $\pi/2$ pulse, contrary to the first one, does not target a specific state starting from a known initial state. It rather realises, in the simplified Bloch sphere picture, a $\SI{90}{\degree}$ rotation around the $J_y$-axis, as depicted in Fig.~\ref{fig:figure1}(c). To optimise this pulse, the following cost function was minimised:
\begin{align}
\label{eq:visibility}
\mathcal{J}^{(2)}&=\max_{t_{\mathrm{hold}}}(1-p_0-p_1) \nonumber\\
\phantom{=}& +\vert 1-\max_{t_{\mathrm{hold}}}(p_0)+\min_{t_{\mathrm{hold}}}(p_0)\vert\nonumber\\
\phantom{=}& +\vert 1-\max_{t_{\mathrm{hold}}}(p_1)+\min_{t_{\mathrm{hold}}}(p_1)\vert
\end{align}
where $p_0$ (resp.~$p_1$) is the ground state (resp.~first excited state) population at the end of the second pulse, and the maximum is taken over $N_h = 15$ different values of the phase accumulation time $t_{\mathrm{hold}}$ for which the numerical optimisation was performed. The first term of equation~\eqref{eq:visibility} minimises the transfer of population to higher energy levels, while the second term (resp.~third term) maximises the amplitude of the oscillation of $p_0$ (resp.~$p_1$). The obtained pulse has a duration of \SI{1.6}{\milli\second}. 

When simulating the whole interferometric sequence, we observe an oscillation of $p_0$ and $p_1$ as a function of $t_{\mathrm{hold}}$, with a periodicity of \SI{0.58}{\milli\second}. The contrast, defined as $\mathcal{C}(p_i) = (\max(p_i)-\min(p_i)) / (\max(p_i) + \min(p_i))$, reaches $\mathcal{C}(p_0) \approx \mathcal{C}(p_1) \approx \SI{97}{\percent}$ in the numerical simulations. As shown in Fig~\ref{fig:figure3}(c), a limited transfer of population to higher excited states on the order $\SI{10}{\percent}$ also takes place. We note that although the second pulse is designed without constraint on the shape of the interferometric fringes, the final fringe evolution is close to a sine function.

% ---- Figure 3 ----

Figure~\ref{fig:figure3}(a) shows the experimentally realised Ramsey signal as obtained by our state analysis (see Supplementary Information). The experimental results are in good agreement with the numerical simulation on the first interferometric fringes. The contrast reaches $\SI{92 \pm 5}{\percent}$, and the Ramsey period measured is $\SI{0.57 \pm 0.02}{\milli\second}$. The fit residuals, interpreted as population in higher excited states and an incoherent fraction, amount to ${\SI{15}\percent} \textendash \SI{25}{\percent}$ depending on $t_{\mathrm{hold}}$.

We point out that the holding times $t_{\mathrm{hold}}$ chosen for the experiment differ from the ones used for the numerical optimisation of the second $\pi/2$ pulse. This indicates that the pulse is valid for all points on the equator of the Bloch sphere. We have investigated this further numerically with other states within the two-modes subspace, but not necessarily lying on the equator of the Bloch sphere. We found that the second $\pi/2$ pulse performs a close-to-unitary operation, similar to a Hadamard gate, with limited leakage to higher excited states.

Looking at longer times $t_{\mathrm{hold}}$ we observe a reduction of contrast, indicating a loss of coherence in the created superposition over time. Fitting an exponentially damped sine to the experimental fringes reveals a damping time constant of \SI{1.6 \pm 0.7}{\milli\second}. This decay is not observed in our 1D GPE simulation (see Fig.~\ref{fig:figure3}(b)).

We investigated three possible mechanisms that could explain the contrast reduction. However, none of them demonstrated decay. (i) Perturbations of the wavefunction could arise from a coupling between the different transverse and longitudinal modes. However, simulations using a 3D GPE solver revealed no such effect. (ii) We evaluated the rate of dephasing~\cite{Castin1997,Garcia-March2012} between the two modes arising from interactions and binomial number fluctuations in each mode and found $R \sim \SI{52}{\milli\radian\per\milli\second}$, hence a dephasing of $\SI{1}{\radian}$ only after $\sim \SI{20}{\milli\second}$, which is too long to account for the observed decay (see Supplementary Information for detail). (iii) Collisional decay of the quantum gas trapped in the excited states would lead to emission of momentum-correlated atom pairs~\cite{Bucker2011}. We do not observe such pair creation in the present experiment, likely due to the lower population in $|1_y\rangle$ and the shorter observation times compared to~\cite{Bucker2011}. 

Furthermore, we observe that the loss of coherence in a situation where the BEC is entirely transferred to the first excited state~\cite{Bucker2011} is slower, on the order of \SI{5}{\milli\second}, which corresponds to the timescale of the correlated atom pairs emission. This suggests that the observed decay is intrinsically related to the specific superposition created by the first pulse of the interferometer. This superposition is a highly excited non-stationary state characterised by non-trivial collective oscillations and may be subjected to damping via scattering.

%----------------------------------------------
%\paragraph{Outlook:}

Ramsey interferometry using motional states introduces a new tool to study out-of-equilibrium evolution of coherent systems at the quantum level~\cite{Scelle2013}.  This may help to shed light onto the mechanisms responsible for the loss of coherence in many-body systems, in particular show the role of interactions. Moreover, fast and coherent manipulation of motional states in a many-body quantum system offers many possibilities that go far beyond Ramsey interferometry. It permits the implementation of general gate operations, the encoding of information into motional states~\cite{Martinez-Garaot2013}, and more generally opens up new perspectives for the use of many-body systems as a viable element in quantum technological applications.

%----------------------------------------------
\section{Acknowledgements}

\begin{acknowledgments}
S.v.F. and R.B. acknowledge the support of the Austrian Science Fund (FWF) through the project CAP (I607-N16). J.-F.S. acknowledges the support of the FWF through his Lise Meitner fellowship (M 1454-N27). T.B. and R.B. acknowledge the support of the Vienna Doctoral Program on Complex Quantum Systems (CoQuS). A.N. acknowledges the support of the excellence cluster 'The Hamburg Centre for Ultrafast Imaging - Structure, Dynamics and Control of Matter at the Atomic Scale' of the Deutsche Forschungsgemeinschaft. T.C. and S.M. acknowledge support from the EC project SIQS and from the DFG via the SFB/TRR21 ``Co.Co.Mat.''. This research was supported by the European STREP project QIBEC (284584), the FWF project SFB FoQuS (SFB F40), the CAP project and the Wittgenstein prize. We are grateful to Thomas Betz, Stephanie Manz, Aur\'elien Perrin, Julian Grond and Ulrich Hohenester for previous work on related projects. We also thank Miguel Angel Garc\'{i}a-March and Wolfgang Rohringer for helpful discussions.
\end{acknowledgments}

%----------------------------------------------
\section{Additional information}

The authors declare no competiting financial interests.

%----------------------------------------------------------------------------------------
%	Additional sections
%----------------------------------------------------------------------------------------

\section{Supplementary information}

\appendix

\section{Trapping potential}

The one-dimensional trapping potential is realised on an atomchip~\cite{Trinker2008} by a radially symmetric Ioffe-Pritchard field modified by radio-frequency dressing~\cite{Schumm2005, Lesanovsky2006,Hofferberth2007}, as explained in detail in Ref.~\cite{Bucker2013}. In the present experiment, the AC current applied has a peak-to-peak amplitude $I_{RF} = \SI{20}{\milli\ampere}$ with detuning $\delta = - \SI{54}{\kilo\hertz}$ with respect to the Larmor frequency near the trap minimum ($\nu_0 = \SI{824}{\kilo\hertz}$). 

In the $y$-direction, where the displacement occurs, the potential is well approximated by the 6th-order polynomial
\begin{equation}
\label{eq:Vy}
V(y) = \alpha_2 y^2 + \alpha_4 y^4 + \alpha_6 y^6 ,
\end{equation}
with $\alpha_2 = h \times \SI{1331}{\hertz}/(2 r_{0,y}^2)$, $\alpha_4 = h \times \SI{62.7}{\hertz}/r_{0,y}^4$ and $\alpha_6 = - h \times \SI{0.63}{\hertz}/r_{0,y}^6$, and where $r_{0,y} = \SI{252}{\nano\meter}$ is the oscillator length in the $y$-direction. The energy differences between the first three single-particle levels of the potential are $ E_{01} = h \times \SI{1.83}{\kilo\hertz}$ and  $ E_{12} = h \times \SI{1.98}{\kilo\hertz}$. In the other directions, the confinement remains essentially harmonic with $\omega_z = 2 \pi \times \SI{2.58}{\kilo\hertz}$. 

In the $z$-direction, it can be described by a quartic polynomial of the form $ V_z = \alpha_2^z z^2 + \alpha_4^z z^4$, with the coefficients $\alpha_2^z = h \times \SI{2516}{\hertz}/(2 r_{0,z}^2)$ and $\alpha_4^z = h \times \SI{17.1}{\hertz}/r_{0,z}^4$ where $r_{0,z} = \SI{212}{\nano\meter}$ is the oscillator length in this direction. This gives a first level spacing $ E_{01}^z = h \times \SI{2.58}{\kilo\hertz}$. 

To create motional states superpositions, we displace the trap minimum along the $y$-direction. This displacement is achieved by modulating the radio-frequency currents with a low-frequency signal. The frequencies of this signal, on the order of a few kilohertz, are much lower than the Larmor frequency of the atoms but higher than the limit for adiabatic displacement of the wavefunction in the transverse potential. This modulation displaces the potential minimum along $y$, following a control trajectory calculated by OCT. The atomic cloud is ``shaked'' by this fast potential displacement.

\section{Optimisation with the CRAB algorithm}

The goal of optimal control is to find the best path in the control parameter space, which is expressed formally as a minimisation of a cost function or performance measure~\cite{Peirce1988, Krotov1996}.

For the optimisation, we describe the system as a condensate wavefunction using an effective one-dimensional GPE along the $y$-axis, with the Hamiltonian
\begin{align}
\label{eq:Hin}
\hat H_{\mathrm{gp}}[\psi,t]=-\frac{\hbar^2}{2m}\frac{\partial^2}{\partial y^2}+V(y - \lambda(t)) + g_\text{y}(N) N\vert\psi(y,t)\vert^2
\end{align}
where $\hbar$ is the reduced Planck constant, $m$ the atomic mass, $N$ the number of atoms and $g_\text{y}(N)$ the effective one-dimensional interaction constant in the $y$-direction~\cite{Salasnich2002}. The minimum of the potential $V$ can be spatially displaced along $y$ by a distance $\lambda(t)$  (see Fig.~\ref{fig:figure1}(a)). 

The CRAB optimisation method expands the control pulse into a (not necessarily orthogonal) basis. Here, the optimisation is carried on 60 Fourier components with their respective amplitudes and phases. Under the action of the control pulse, the wavefunction undergoes a transformation that is computed numerically using split-step analysis~\cite{Agrawal2001}. The wavefunctions of the different motional states are the stationary solutions of the GPE and were obtained numerically by imaginary time propagation~\cite{Auer2001}.

\section{Extraction of populations}

Because the effect of interactions in the $y$-direction is already negligible after a very short time-of-flight (less than $\SI{1}{\milli\second}$), we assume a ballistic expansion. The density distribution imaged after time-of-flight is then homotetic to the momentum distribution. 

To simulate this distribution, we calculate the evolution of the 1D GPE in the static potential $V(y)$ starting from a given initial superpositon of $k$ states
\begin{equation}
\label{eq:Psi_start}
| \psi_{\text{initial}} \rangle = \sum_k \sqrt{p_k} e^{i \theta_k} | k_y \rangle
% \sqrt{p_0} | 0_y \rangle + \sqrt{p_1} | 1_y \rangle + \sqrt{p_2} | 2_y \rangle,
\end{equation}
where $k \in \{0,1,2\}$, corresponding to the three lowest-lying states in the $y$-direction. We compute the momentum distribution and compare its evolution to the experimental densities after time-of-flight. 

To recover the wavefunction superposition after the control pulse from experimental data, we fit the data with a time-dependent momentum density along $y$. The numerical momentum density is obtained by Fourier transform of an in-trap GPE simulation. Experimentally, we can access the atomic density after $\SI{46}{\milli\second}$ time-of-flight. The fast transverse expansion of the cloud due to high confinement causes the atomic interactions to become rapidly negligible, hence the expansion can be considered ballistic. In the limit of infinite expansion time, the in-trap momentum distribution and the density after time of flight are strictly equivalent. Here, the time of flight is sufficiently long to make this assumption. If we express the momenta as wave numbers $k_y$, a distance $\delta y$ in the experimental image then corresponds to $\delta k_y = \alpha \delta y$ with $\alpha = m / \hbar t_{TOF} \approx \SI{0.03}{\per\square\micro\meter}$. The simulated momentum distribution is slightly rescaled on the $k$-axis and corrected for imaging broadening, then sampled to match the experimental sampling time $t = \SI{0.05}{\milli\second}$. 

We simulate the momentum distribution density with, as fit parameters, the atomic fractions in the first three motional states $p_0$, $p_1$ and $p_2$, as well as the relative phases between those states $\theta_{01}$ and $\theta_{12}$. We chose to restrict the model to a three states superposition here. First, multi-mode simulations show that the main features of the experimental data can be reproduced by a 3-mode description similar to Ref.~\cite{Garcia-March2012}. Second, this assumption is justified by the fact that adding more states does not improve nor modify much the output of the fit. 

From there, the similarity between simulated and experimental densities is evaluated by substracting both and calculate the residual. To obtain the combination of parameters most likely to have generated the observed momentum distribution density, we use a simplex regression method that searches the smallest possible residual and gives their corresponding best values for the fit parameters. Once these parameters are obtained, we look for the uncertainty of the fit by estimating the variances and co-variances of the different parameters and deduce the confidence intervals of the fit. 

We note that these fits are based on Gross-Pitaevskii simulations, which represent a unitary evolution for a mean-field description of a system at zero temperature. Although this model describes the main features of our data very well, some discrepancy between the model and the experiment (e.g. many-body or finite temperature effects) may have systematic effects on the estimation of the fidelity. It is nevertheless unlikely that these discrepancies have a qualitative effect on the interferometer output.

\section{Phase diffusion}

We estimated the rate of many-body dephasing that could arise from number fluctuations in the ground and excited states. We followed the approach of Ref.~\cite{Garcia-March2012}, and, assuming weak interaction, approximated the field operator $\hat \psi$ describing the BEC by:
\begin{equation}
\hat \psi(y) \simeq \hat a_0 \phi_0(y) + \hat a_1 \phi_1(y) . \label{eq:Ansatz_psi}
\end{equation}
Here the $\phi_i$'s are the two lower-lying eigenstates of the non-interacting part of the Hamiltonian (taken to be real and normalized to $\int |\phi_i|^2 {\rm d}y = 1$), the $a_i$'s are annihilation operators associated with the modes and a 1D geometry along the $y$-axis was assumed for simplicity. From the full many-body Hamiltonian describing the condensate and equation~\eqref{eq:Ansatz_psi}, we obtain the following effective 2-mode Hamiltonian:

\begin{equation}
\hat H_\text{2m} = \Delta E \, {\hat J}_z + U {\hat J}_z^2 + 4 U_{01} {\hat J}_x^2 , \label{eq:hamiltonian} 
\end{equation}
with
\begin{align}
\Delta E &= E_{01} - (N-1)(U_{00} - U_{11}) , \\
U &= U_{00} + U_{11} - 2 U_{01} , \label{eq:interaction} \\
\text{and} \quad U_{ij} &= \frac{1}{2} g_\text{1D} \int |\phi_i|^2 |\phi_j|^2 {\rm d}y ,
\end{align}

where we used the usual spin representation for the many-body two level system by introducing the operators ${\hat J}_x = (\hat a_0 \hat a_1^\dagger + \hat a_0^\dagger \hat a_1)/2$, ${\hat J}_y = (\hat a_0 \hat a_1^\dagger - \hat a_0^\dagger \hat a_1)/2i$ and ${\hat J}_z = (\hat a_1^\dagger \hat a_1 - \hat a_0^\dagger \hat a_0)/2$, which satisfy angular momentum commutation relations. 
This Hamiltonian resembles the bosonic Josephson Hamiltonian in the presence of an energy offset between the two modes, here given by the difference of chemical potential between the ground and first excited states (first term $\propto {\hat J}_z$). The second term $\propto {\hat J}_z^2$, which comes from interactions, is responsible for ``phase diffusion'' (dephasing). It leads to squeezing at short times~\cite{Kitagawa1993}, generation of strongly non-classical states~\cite{Piazza2008} and a loss of coherence at longer times~\cite{Javanainen1997, Castin1997}. The third term is generally not present in bosonic Josephson junctions.

In the second term of equation~\eqref{eq:hamiltonian}, it is apparent that phase diffusion is reduced compared to e.g.\ the case of a double well system~\cite{Berrada2013}, as the modes have a significant spatial overlap. This is similar to the case of a spinor condensate in which two spin states share the same external wavefunction and have similar scattering lengths~\cite{Steel1998, Gross2010}. We can evaluate the phase diffusion rate if we assume e.g.\ a binomial distribution of the atoms in each mode (i.e.\ $\Delta {\hat J}_z = \sqrt{N}/2$), which is a fair assumption if the first $\pi/2$ pulse is performed quickly compared to the other energy scales (in particular compared to interactions that may induce squeezing). It is then given by~\cite{Javanainen1997, Castin1997}:

\begin{equation}
R = \frac{2\Delta {\hat J}_z U}{\hbar} .
\end{equation}

We computed the two wavefunctions $\phi_0$ and $\phi_1$ in the trapping potential $V_y$ and obtained the energies $U_{00}/h = 0.34$\,Hz, $U_{11}/h = 0.26$\,Hz, and $U_{01}/h = 0.15$\,Hz. This yields $U/h = 0.31$\,Hz, and a phase diffusion rate $R = 52$\,mrad/ms. This rate increases with atom number fluctuations and can become significant if the fluctuations are much stronger than in the binomial case ($\Delta n \gtrsim 20 \sqrt{N}$, $n$ being the population difference between the modes). This could be the case, as both modes overlap.

%----------------------------------------------
%\section{References}

\end{document}